\documentclass{pic2012}
\def\runauthor{R. Leitner}
\def\shorttitle{The Experimental Status of $\theta_{13}$ - reactor neutrino experiments}
\begin{document}

\title{The Experimental Status of $\theta_{13}$ from the Point of View of the Electron (Anti-) Neutrino Disappearance Experiments \\
}

\author{Rupert Leitner\footnote{\scriptsize{The Daya Bay collaboration}}} 

\address{
Faculty of Mathematics and Physics, Charles University, Prague\\
V Hole\v{s}ovi\v{c}k\'{a}ch 2, CZ 180 00 Praha, Czech Republic\\
E-mail: Rupert.Leitner@mff.cuni.cz}

\maketitle

\abstracts{A non zero, surprisingly large value of the third mixing angle $\theta_{13}$ has been measured in reactor neutrino experiments. Currently the  most precise measurement of $\sin^2 2\theta_{13}$ has been performed by the Daya Bay experiment $\sin^22\theta_{13}=0.089\pm 0.010({\rm stat.})\pm0.005({\rm syst.})$ (7.7 $\sigma$ significance of $\sin^22\theta_{13} > 0$ ),the RENO experiment has measured the value $\sin^2 2\theta_{13} = 0.113 \pm 0.013(\rm stat.) \pm 0.019(\rm syst.)$ (4.9 $\sigma$ significance) and the Double Chooz experiment $\sin^2 2\theta_{13} = 0.109 \pm 0.030(\rm stat.) \pm 0.025(\rm syst.)$ (2.9 $\sigma$ significance). These results are extremely important for future searches of violation of combined CP parity in lepton sector of the Standard model. 
}

\section{Neutrino mixing and oscillations} 
Neutrino flavor eigenstates $\nu_{f}$ originating in weak decays together with charged leptons of three known flavors $f=e,\mu,\tau$ are super-positions of three mass eigenstates $\nu_{i}$ $i=1,2,3$ with masses $m_{1}, m_{2}, m_{3}$. The mixing is described by $3x3$ unitary Pontecorvo-Maki-Nakagawa-Sakata (PMNS) matrix $U$ with elements
\begin{eqnarray}
 U_{fi} \equiv \langle \nu_{f} |\nu_{i} \rangle \nonumber
\end{eqnarray}
Using this mixing matrix one can write flavor eigenstates as super-positions of mass eigenstates as follows: 
\begin{eqnarray}
|\nu_{f} \rangle = \sum_{i} |\nu_{i} \rangle \langle \nu_{i} |\nu_{f} \rangle = \sum_{i} U^{*}_{fi} |\nu_{i} \rangle .  \nonumber 
\end{eqnarray}

The CPT invariance implies the equivalence of masses of neutrinos and anti-neutrinos and relates the mixing matrices for neutrinos ($U$) and anti-neutrinos ($\bar{U}$):  
\begin{eqnarray}
\bar{U}_{fi} \equiv  \langle \bar{\nu}_{f} |\bar{\nu}_{i} \rangle =(CPT)= \langle \nu_{i} |\nu_{f} \rangle = U^{*}_{fi} .\nonumber     
\end{eqnarray}
The mixing of anti-neutrinos flavor and mass eigenstates therefore obeys the relation:
\begin{eqnarray}
|\bar{\nu}_{f} \rangle = \sum_{i} U_{fi} | \bar{\nu}_{i} \rangle \nonumber
\end{eqnarray}

Experimentally firmly established phenomenon of oscillations of neutrino flavors can be explained by non-diagonal mixing matrix and different masses of neutrinos mass  eigenstates. In general the 3x3 unitary matrix contains 9 free parameters. For Dirac neutrinos one can reduce the number of physical parameters to 4 by re-phasing five out of six lepton fields. Canonical representation of the mixing matrix is the ordered product of three rotations with angles $\theta_{12}$, $\theta_{13}$, $\theta_{23}$ and one CP violating phase $\delta$. In case of Majorana origin of neutrinos the mixing matrix contains another two so called Majorana phases $\alpha_{1,2}$. Oscillations of neutrino flavors do not depend on Majorana phases; the relevant for oscillations part of the mixing matrix is: 

\begin{eqnarray}
 U_{fi} = 
\left( \begin{tabular}{ccc}
         $1$ & $0$ & $0$ \\
         $0$ & $c_{23}$ & $s_{23}$ \\
         $0$ & $-s_{23}$ & $c_{23}$ 
         \end{tabular}
\right)
\left( \begin{tabular}{ccc}
         $c_{13}$ & $0$ & $s_{13} e^{i\delta} $  \\
         $0$ & $1$ & $0$ \\
         $-s_{13} e^{-i\delta}$ & $0$ & $c_{13}$ 
         \end{tabular}
\right)
\left( \begin{tabular}{ccc}
         $c_{12}$ & $s_{12}$ & $0$ \\
         $-s_{12}$ & $c_{12}$ & $0$ \\
         $0$ & $0$ & $1$ 
         \end{tabular}
\right) \nonumber 
\end{eqnarray}
where $s_{ij}$ and $c_{ij}$ denote $\sin{\theta_{ij}}$ and $\cos{\theta_{ij}}$ respectively.  

With an approximation of $\Delta m^{2}_{31} \equiv m^{2}_{3} - m^{2}_{1} \cong  m^{2}_{3} - m^{2}_{2} \equiv \Delta m^{2}_{32}$ the survival probability for electron anti-neutrino with energy $E$ at the distance $x$ from the source is given by the formula: 
\begin{eqnarray}
P_{\bar{\nu}_{e}\rightarrow \bar{\nu}_{e}} \left( \frac{x}{E} \right) = 1 - \sin^{2} \left( 2\theta_{13} \right) \sin^{2} \left(  \frac{\Delta m^{2}_{31} x }{4\hbar c E} \right) - \cos^{4} \left( \theta_{13} \right) \sin^{2} \left( 2\theta_{12}  \right) \sin^{2} \left(  \frac{\Delta m^{2}_{21} x }{4\hbar c E} \right) 
\nonumber
\end{eqnarray}

Values of two mass square differences and mixing angles $\theta_{12}$ and $\theta_{23}$ have been measured in following experiments:
\begin{itemize}
\item 
experiments with atmospheric and accelerator neutrinos: \\
$|\Delta m^{2}_{31} | = | m^{2}_{3} - m^{2}_{1} | \cong |\Delta m^{2}_{32} | = | m^{2}_{3} - m^{2}_{2} | \cong \left(48.2~meV\right)^{2}$ and $\theta_{23} \cong 45^{o}$. 
\item
experiments with reactor and Sun neutrinos: \\
$\Delta m^{2}_{21} = m^{2}_{2} - m^{2}_{1} \cong \left(8.7~meV\right)^{2}$ and 
$\theta_{12} \cong 34^{o}$.
\end{itemize}
Because $|\Delta m^{2}_{31} | \cong 30 \cdot \Delta m^{2}_{21}$ there are two oscillations with 30 times different distances to the 1st oscillation minimum $L_{1^{st}min}$:

\begin{tabular}{ll}
 $|\Delta m^{2}_{31}| \cong |\Delta m^{2}_{32}| $  & $L_{1^{st}min} = 0.5 km/MeV =   500 km/GeV$ \\
 $\Delta m^{2}_{21}$ & $L_{1^{st}min} =  15 km/MeV = 15000 km/GeV$ 
\end{tabular} 

The aim of three reactor neutrino experiments Daya Bay \cite{DYB}, Double Chooz \cite{DC} and RENO \cite{RENO} was to measure not observed before disappearance of reactor electron anti-neutrinos at distances of 0.5 $km/MeV$. Such observation would be a discovery of non zero value of the mixing angle $\theta_{13}$. 
An alternative way to measure the value of $\theta_{13}$ is via the observation of the appearance of electron (anti-)neutrinos in accelerator muon (anti-)neutrinos at distances 
500~$km/GeV$. 
  
\section{Detection of reactor anti-neutrinos}
Nuclear power reactors are very powerful sources of electron anti-neutrinos. Fission products are neutron rich isotopes that are transformed to stable nuclei via series of $\beta^{-}$ decays. In average $\approx$ 6 electron anti-neutrinos follow each fission producing the flux of $\approx 2 \cdot 10^{20}$ of $\bar{\nu}_{e}$ per second per $1~GW_{th}$ of reactor thermal power. 
Energy spectrum of anti-neutrinos decreases with the energy and extends up to $\approx 10~MeV$. Anti-neutrinos with energies $E_{\nu} \geq 1.8~MeV$ can be detected by charged current interactions (so called inverse beta decay IBD):

\begin{eqnarray}
\bar{\nu}_{e} + p \rightarrow e^{+} + n \nonumber
\end{eqnarray}

on target protons of liquid scintillator.
Detected anti-neutrinos energy spectrum is the product of original anti-neutrino flux and neutrino energy dependent IBD cross section. Detected spectrum has a typical shape with a maximum at $E_{\nu} \approx 4~MeV$. 

\subsection{Detection of positrons}
Positron losses its kinetic energy $T_{e}$ by ionization and annihilates with an electron producing pair of $m_{e}=511~keV$ gammas. This so called prompt signal $E_{p}$ 
\begin{eqnarray}
E_{p} & = &T_{e}+2m_{e} \nonumber 
\end{eqnarray}
is related to neutrino energy as follows:
\begin{eqnarray}
E_{\nu} & \cong & E_{p}+ 0.8~MeV  \nonumber
\end{eqnarray}

\subsection{Detection of neutrons}
Kinetic energy of IBD neutrons is $\approx$ 4 $keV$ and 40 $keV$ for neutrinos with energies 4 $MeV$ and 10 $MeV$ respectively. Neutrons are thermalized by elastic collisions with Hydrogen and Carbon nuclei of liquid scintillator and then captured. In order to distinguish neutron signal from backgrounds, liquid scintillator is doped by Gadolinium. Two isotopes 155 and 157 of natural Gd have huge thermal neutrons capture cross sections of 61 and 254 k barns respectively. Neutron capture on these Gd isotopes is followed by the emission of several gammas with total energies of 8.54 $MeV$ and 7.94 $MeV$ respectively.    
Because the natural abundances of $^{155}Gd$ and $^{157}Gd$ are high (14.8$\%$ and 15.7$\%$) the addition of a small amount of ~0.1$\%$ of Gd is enough to ensure a capture of more than 80$\%$ of neutrons on Gd (and less then 20$\%$ on H of liquid scintillator). 
Time difference $\Delta t$ between prompt signal from positron and delayed signal from neutron follows an exponential distribution $e^{-\Delta t/\langle \tau \rangle}$ with the value of capture time $\langle \tau \rangle \cong 29 \mu s $ inversely proportional to Gd concentration. 

\section{Reactor neutrino experiments}

\begin{figure}[tbh]
\vspace*{7cm}
\begin{center}
\includegraphics{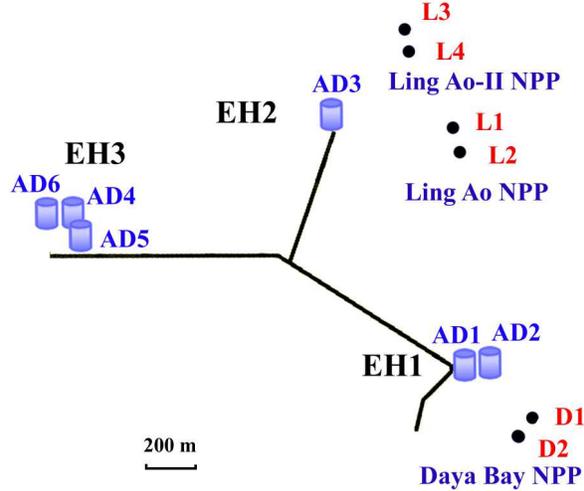}
\caption[*]{ Layout of the Daya Bay experiment.
The dots represent reactor cores, labelled as D1, D2, L1, L2, L3 and L4.
Six anti-neutrino detectors (ADs) were installed in three experimental halls (EHs).The figure is taken from \cite{DYB}. \label{fig:layout}}
\end{center}
\end{figure}

Number of detected events is proportional to the product of the detector mass and power of nuclear reactors. In addition optimal sensitivity requires to place far detector(s) at the distance close to 2 $km$ where the maximal disappearance is expected for neutrinos around peak energy of 4 $MeV$. 
Correlated systematic errors are cancelled and uncorrelated reactor uncertainties are minimized in a configuration with identical near detector(s) \cite{MIK} placed closer to reactors where oscillations effects are small. The background can be minimized by large overburden.  The near-far arrangement of Daya Bay anti-neutrino detectors (ADs), is shown on Fig.~\ref{fig:layout}, relevant parameters for all the three experiments are summarized in the following table:\\

\begin{center}
\begin{footnotesize}
\begin{tabular}{|l|c|cc|cc|cc|}
\hline
 & Reactor Power & \multicolumn{2}{c|}{Detector mass [t]} & \multicolumn{2}{c|}{Baseline [m]} & \multicolumn{2}{c|}{Overburden [mwe]} \\
 Experiment& [GW$_{th}$] & Near & Far & Near & Far & Near & Far \\
\hline
Daya Bay & 17.4 & 40, 20$^{*}$ & 60$^{*}$ & 470, 576 & 1650 & 250, 265 & 860 \\
\hline
Double Chooz & 8.5 & -$^{**}$ & 8.2 & 400 & 1050 & 120 & 300 \\
\hline
RENO & 16.5 & 16.5 & 16.5 & 409 & 1444 & 120 & 450 \\
\hline
\end{tabular}
\end{footnotesize}
\end{center}

Table shows the configurations at the time of the PIC2012 conference. \\
$^{*}$ Since October 2012 the Daya Bay detector has been completed to 40+40 t of near detectors and 80 t of far detectors. \\
$^{**}$ The Double Chooz experiment currently operates with far detector only.

\subsection{Anti-neutrino detectors}

\begin{figure}[ht]
\vspace*{6.0cm}
\begin{center}
\includegraphics{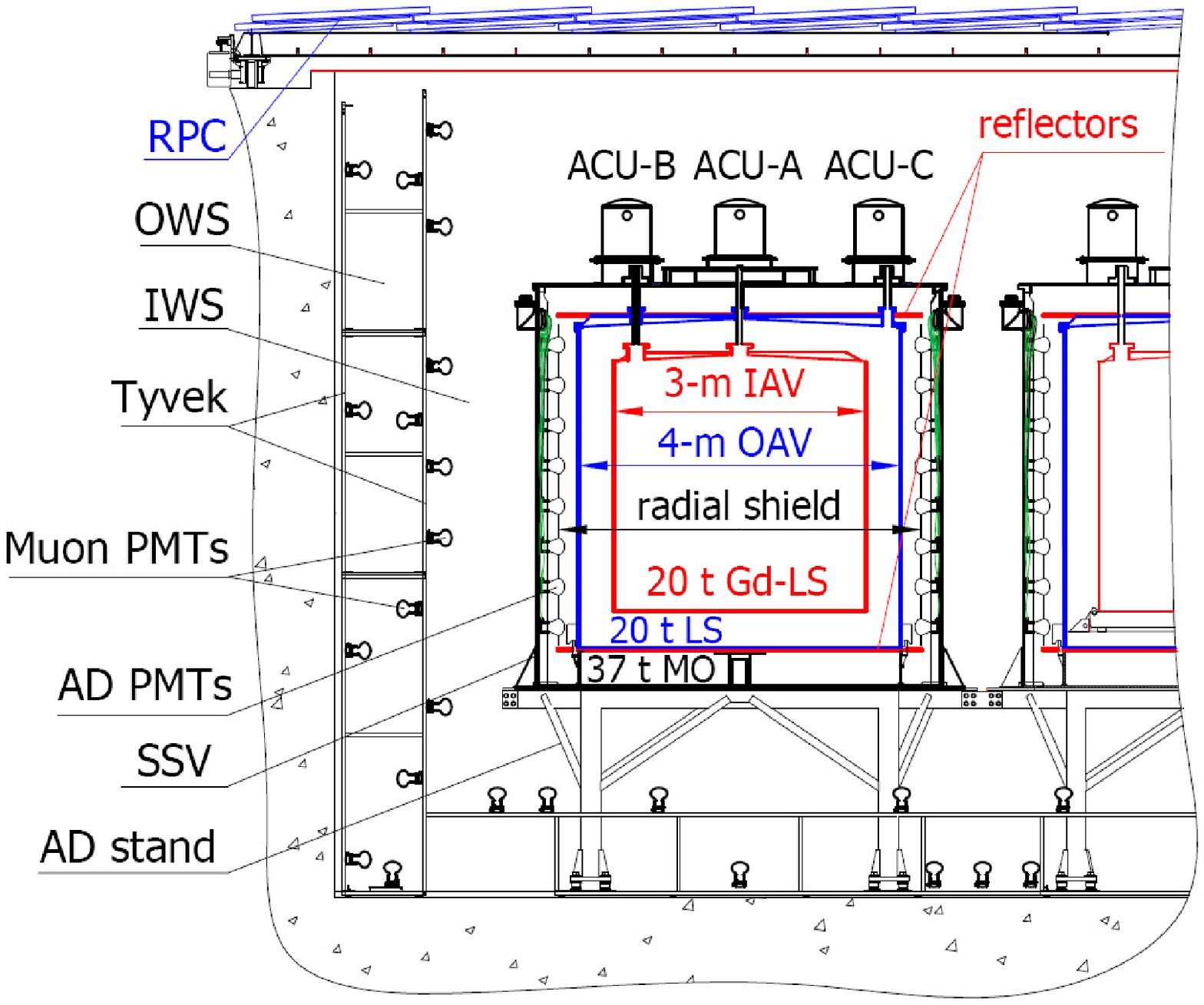}
\includegraphics{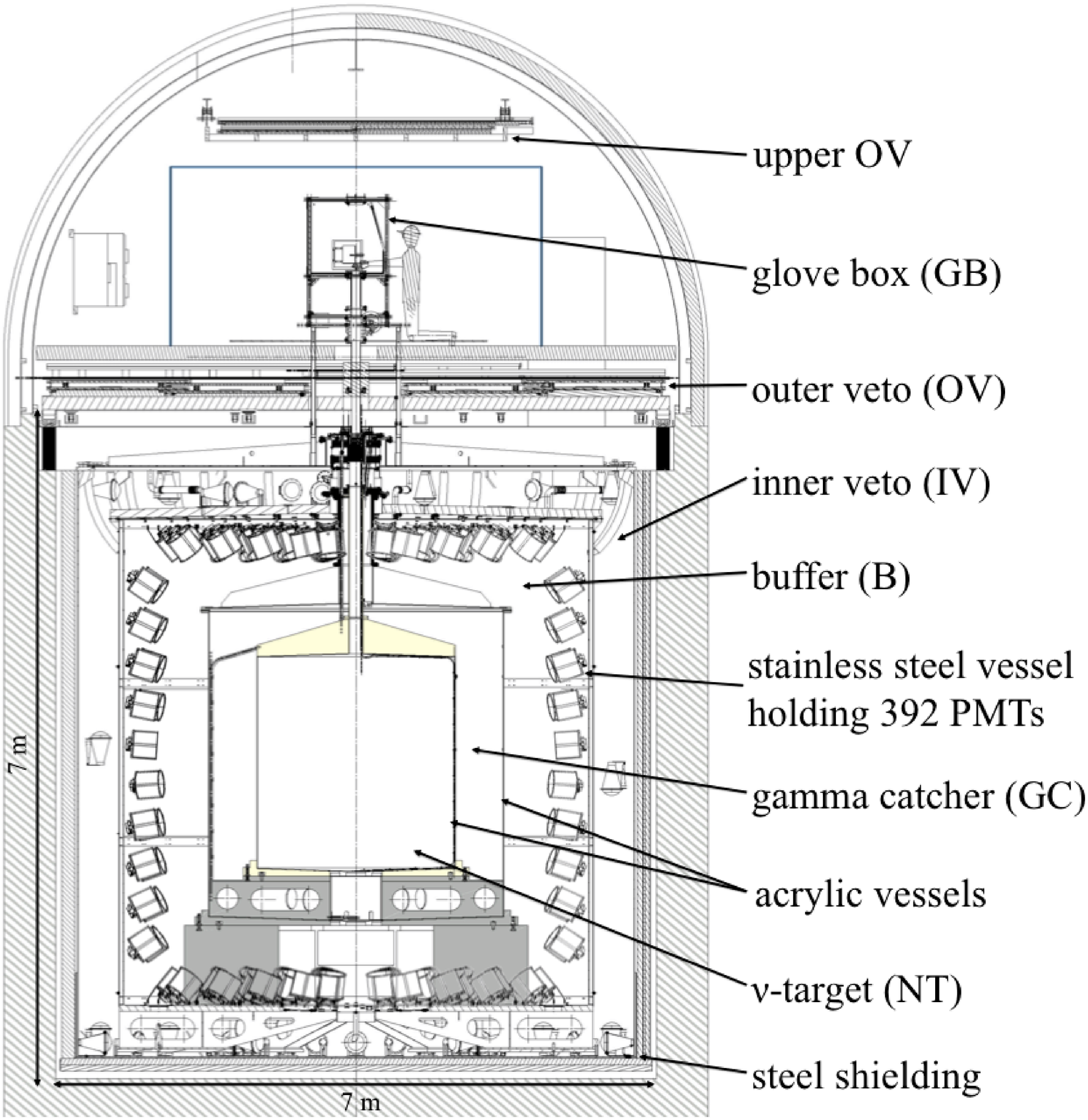}
\caption[*]{ Schematic diagram of the Daya Bay (left picture) and Double Chooz (right picture) detectors. Pictures are taken from Ref.\cite{DYB,DC2}. \label{fig:det} }
\end{center}
\end{figure}

Each of three near and three far Daya Bay anti-neutrino detectors \cite{ad12} is the stainless steel vessel with three nested cylindrical volumes separated by two concentric acrylic vessels (see Figure~\ref{fig:det}). The innermost volume serves as the target and is filled with 20 t of Gd loaded liquid scintillator (Gd-LS). In order to fully contain gammas from positron annihilation and Gd de-excitation the central volume is surrounded by gamma catcher of 21t of Gd free liquid scintillator (LS) in outer acrylic vessel. Outermost volume of 37 t of mineral oil serves as shielding of inner volumes from radiation originating from the walls of stainless steel vessel and photomultipliers. To increase the light yield and improve the homogeneity reflector panels are installed at the top and bottom of the outer acrylic vessel.  Scintillating light is detected by 192 8-inch PMTs placed along the circumference of the stainless steel vessel. 
The Daya Bay detectors are submerged in instrumented water pools such that at least 2.5 m of water is surrounding each detector. Purified water passively shields detectors from the ambient and cosmic rays induced radioactivity and actively detects Cherenkov light produced by cosmic muons. In addition, an array of resistive plate chamber (RPC) modules is placed on top of the pools for increased background rejection.

Target mass of the Double Chooz detector (see Figure~\ref{fig:det}) is 8.5 t (10.3 m$^{3}$) of Gd-LS, surrounded by 22 m$^{3}$ of LS gamma catcher and by 110 m$^{3}$ of mineral oil buffer. The detector is instrumented by 390 10-inch PMTs. The Double Chooz detector is submerged in 90 m$^{3}$ of liquid scintillator active inner veto and it is covered and surrounded by 15 cm thick shield of demagnetized steel. From top it is covered by two active outer veto detectors made out of scintillator strips.        

The RENO experiment operates one near and one far detector each with the target mass of 16 t surrounded by 30 t of gamma catcher and 55 t of mineral oil. The detectors are instrumented with 354 10-inch PMTs and submerged in 350 t of active purified water veto system.

\subsection{Detector calibration}

Detectors of all three experiments are equipped with calibration systems. Daya Bay detectors have three automated calibration units ACU A, B and C (see Figure~\ref{fig:det}) that insert calibration sources along vertical direction in the centre of the detector, close to the edge of target volume and in the middle of gamma catcher; Double Chooz calibration sources are lowered from a glove box (see Figure~\ref{fig:det}) at the detector top; RENO detectors are using three dimensional and one dimensional calibration system. 

Double Chooz and RENO detectors are calibrated by gammas from $^{137}Cs$ radioactive source:
\begin{eqnarray}
^{137}Cs & \rightarrow & ^{137m}Ba  + e^{-} + \bar{\nu}_{e} \nonumber \\
        &              & ^{137m}Ba   \rightarrow  ^{137}Ba + \gamma (661.7~keV)  \nonumber
\end{eqnarray}

The calibration of all three experiments near the IBD threshold is done with $^{68}Ge$ that decays by electron capture to $\beta ^{+}$ emitter $^{68}Ga$ following the decay scheme:
\begin{eqnarray}
e^{-} + ^{68}Ge & \rightarrow & ^{68}Ga + \nu_{e} \nonumber \\
        &             & ^{68}Ga  \rightarrow  ^{68}Zn + e^{+} + \nu_{e} \nonumber ;
\end{eqnarray}

detectors response to pair of 511 $keV$ gammas from positron annihilation in the source capsule is reconstructed. 

Near the peak of detected anti-neutrino energy spectrum detectors of all three experiments are calibrated with $^{60}Co$ radioactive source: 
\begin{eqnarray}
^{137}Co & \rightarrow & ^{60}Ni^{*}  + e^{-} + \bar{\nu}_{e} \nonumber \\
        &              & ^{60}Ni^{*}  \rightarrow  ^{60}Ni + \gamma (1173.2~keV)  + \gamma (1332.5~keV) \nonumber
\end{eqnarray}

Neutrons from $^{241}Am-^{13}C$ source (Daya Bay) and from spontaneous fissions of $^{252}Cf$ (Double Chooz and RENO) are used to calibrate detectors response to gammas from neutron capture on H and Gd.
The detector's response is also monitored by neutrons produced by cosmic rays. 

In addition the PMT timing, gain and relative quantum efficiencies are calibrated using Light-Emitting Diodes (LED) light (Daya Bay and Double Chooz) and Laser light (RENO). 

Energy resolution of Daya Bay detectors has been measured to satisfy the relation $\sigma / E = 7.5 \% / \sqrt{E[MeV]} + 0.9 \%$; energy resolution of RENO follows the relation $\sigma / E = 5.9 \% / \sqrt{E[MeV]} + 1.1 \%$. 
Values of statistical terms of $7.5 \%$ and $5.9 \%$ are in good agreement with the yield of 160 photoelectrons/MeV and 250 photoelectrons/MeV measured by Daya Bay and RENO detectors respectively.

\subsection{IBD candidates selection}
Daya Bay.
Minimum value of the  IBD signal at the threshold is 1.022 $MeV$. Due to energy resolution and possible inhomogeneity of the response, prompt signal is selected in the region $0.7~MeV < E_{p} < 12~MeV$. Delayed signal $6~MeV < E_{d} < 12~MeV$ from the neutron capture on Gd is searched in time window $\Delta t \equiv t_{d}-t_{p} = 1 \div 200~\mu s$ following the prompt signal. In addition there should be no signal exceeding 0.7 MeV in $200 \mu s$ time windows preceding and following the prompt signal and the delayed signal respectively (so called multiplicity cut). 

In order to suppress cosmogenic background Daya Bay detectors are vetoed for 0.6 $m s$, 1 $m s$ and 1 $s$ after so called pool muon, non-showering AD muon and showering AD muon signals respectively. 

Selection criteria for all three experiments are compared in following table. :

\begin{center}
\begin{tabular}{l|c|c|c|l|c}
\hline
Detector      & $E_{p}$      & $E_{d}$ & $t_{d}-t_{p}$ & Multiplicity & Veto      \\
              & MeV          &  MeV    &   $\mu s$     &   $\mu s$    &  $m s$     \\
\hline
Daya Bay      & (0.7; 12)    & (6; 12) & (1; 200)      & 200~/~200    & 0.6~/~1~/~1000 \\
Double Chooz  & (0.7; 12.2)  & (6; 12) & (2; 100)      & 100~/~400    & 500$^{*}$ \\
RENO          & (0.7; 12)    & (6; 12) & (2; 100)      & 100~/~-      & 1~/~10$^{**}$ \\
\hline
\end{tabular}
\end{center}
Two values for multiplicity cut denotes the time window preceding the prompt signal and following the delayed signal respectively.\\
$^{*}$ Double Chooz detector is vetoed for 500 $m s$ after cosmic muon signal $>$600 $MeV$. \\
$^{**}$ RENO detectors are vetoed for 1 $m s$ after AD muon signal larger than 70 $MeV$ (or $>$20  $MeV$ if that coincides with large water pool signal) and for 10 $m s$ after cosmic muon signal $>$1.5 $GeV$.

Total number of IBD candidates collected by each experiment is summarized in the following table:
\begin{center}
\begin{tabular}{|l|c|r|}
\hline
Detector & Near & \multicolumn{1}{c|}{Far} \\
\hline
Daya Bay & 205 308 & 28 909 \\
Double Chooz & - & 8 249 \\
RENO & 154 088 & 17 102 \\
\hline
\end{tabular}
\end{center}

\begin{figure}[ht]
\vspace*{4.5cm}
\begin{center}
\includegraphics{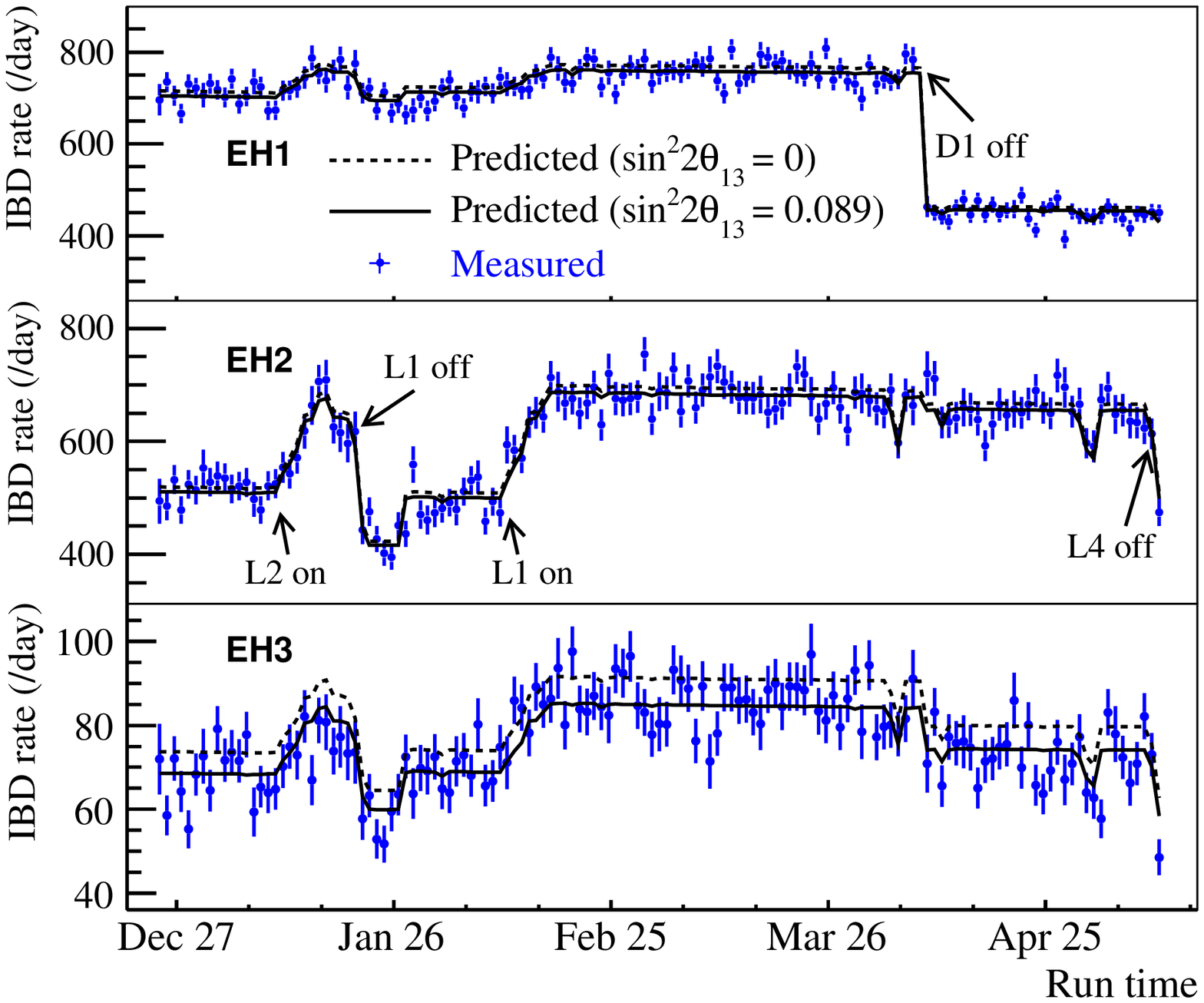}
\includegraphics{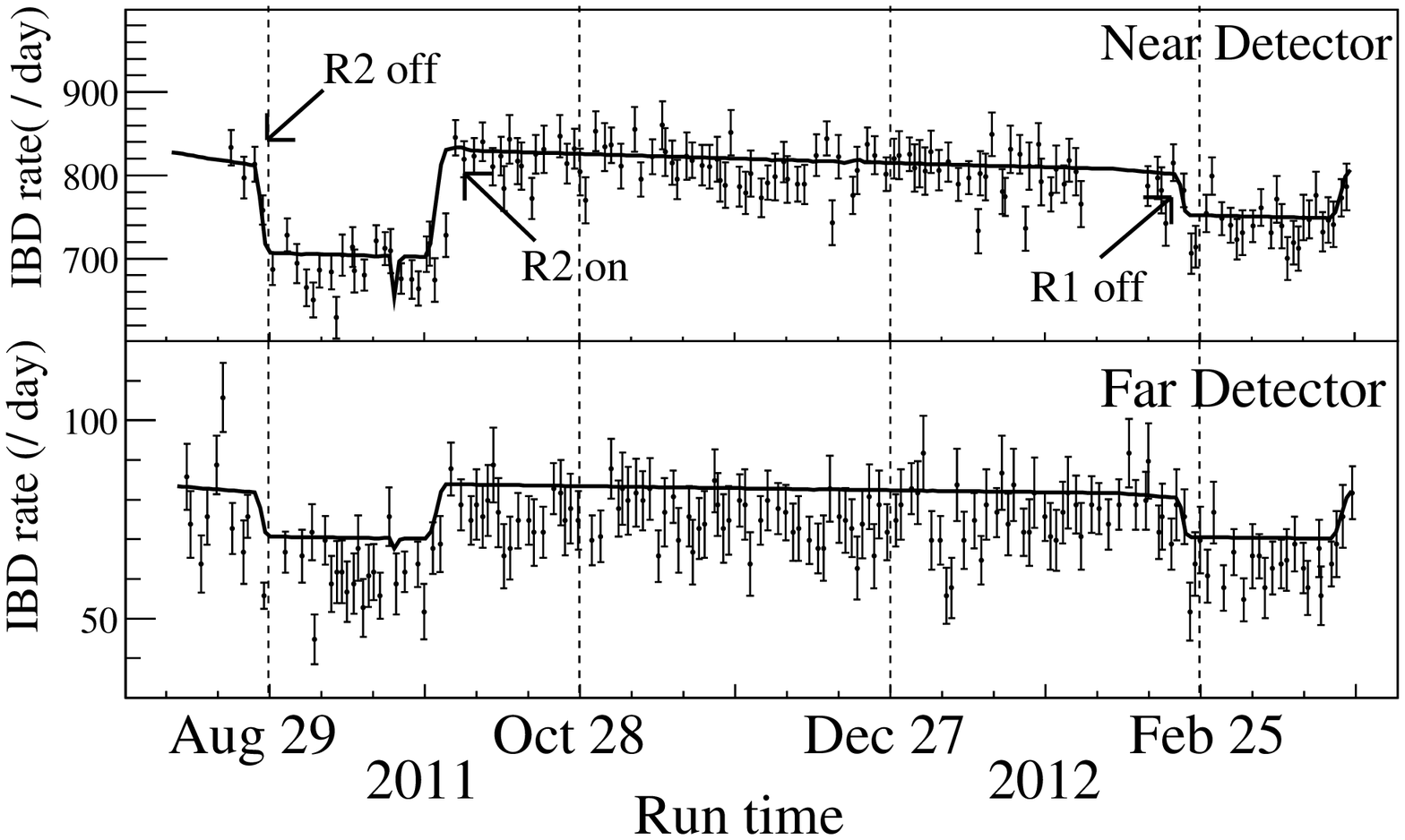}
\caption[*]{ The correlation of daily IBD rates and reactor powers. Left plot shows Daya Bay daily IBD rates per detector in three experimental halls (near EH1 and EH2 and far EH3). Right plot shows RENO daily IBD rates measured in near and far detector. Figures are taken from Ref.\cite{DYB,RENO}.
 \label{fig:dailyrates}}
\end{center}
\end{figure}

The correlation of daily IBD rates per detector and reactor powers for Daya Bay and RENO are shown on Figure~\ref{fig:dailyrates}. 

\subsection{Systematic uncertainties}

The values of systematic uncertainties for all three experiments are summarized in the Table~\ref{tab:syst}. 

\begin{table*}[hb]
\begin{center}
\caption{Comparison of systematic uncertainties. All values are in $\%$. \label{tab:syst}  }
\begin{tabular}{|l|c|c|c|c|c|c|}
\hline
        & \multicolumn{2}{c|}{Detector related }  & \multicolumn{2}{c|}{Reactor related} & \multicolumn{2}{c|}{$\frac{Background}{Signal}$ } \\
Detector & Corr & Uncorr & Corr & Uncorr & Near & Far \\
\hline
Daya Bay     & 1.9 & 0.2 & 3.0 & 0.8 &  1.9 $\pm$ 0.2 & 4.7 $\pm$ 0.35\\
RENO         & 1.5 & 0.2 & 2.0 & 0.9 & 2.8 $\pm$ 0.8  & 5.8 $\pm$ 1.1 \\
Double Chooz &  \multicolumn{2}{c|}{1.0} &  \multicolumn{2}{c|}{1.7} &   & 5.5 $\pm$ 1.6 \\
\hline
\end{tabular}
\end{center}
\end{table*}

For Daya Bay and RENO experiments with near-far configuration of detectors only uncorrelated systematic errors influence the result. Dominant contributions to detector related uncorrelated uncertainties are uncertainties of delayed energy cut efficiency and of fraction of neutron captures on Gd. Due to near-far detector configurations, influences of uncorrelated reactor based uncertainties to measured ratio of far to near IBD rates are highly reduced (by a factor of $\approx$ 20 for the Daya Bay).  

\subsection{Backgrounds}
The signal of IBD events can be mimic by three major types of background events. 
\begin{itemize}
\item Accidental background are coincidences of prompt and neutron-like signals not caused by IBD events. This background is estimated using data and it is concentrated at low energies. 
\item Fast neutrons background. Energetic neutrons created in photonuclear interactions of cosmic muons could mimic IBD by recoiling off a proton before being captured on Gd.  The number of background events can be estimated by the extrapolation of the background shape at energies above 12 MeV.
\item Very neutron rich exotic nuclei created in interactions of cosmic muons. These are mostly $_{3}^{9}Li$ and less frequently also $_{2}^{8}He$.
Large fraction of $_{3}^{9}Li$ and $_{2}^{8}He$ $\beta^{-}$ decays is accompanied by free neutron:   
\begin{eqnarray}
 _{3}^{9}Li & \rightarrow & _{4}^{8}Be + n + e^{-} + \bar{\nu_{e}} ~~~~~ Q=13.607~MeV \nonumber \\
 & & _{4}^{8}Be  \rightarrow    2 \alpha  \nonumber \\
       \nonumber  \\
 _{2}^{8}He & \rightarrow & _{3}^{7}Li + n + e^{-} + \bar{\nu_{e}} ~~~~~ Q=10.651~MeV \nonumber 
\end{eqnarray}
Due to large Q value $\beta^{-}-n$ decays of these nuclei mimic IBD candidates spanning the whole spectrum and because of their long half-lives of 178 and 119~$ms$ detectors shall be vetoed for long times after showering cosmic muon signals. 
\end{itemize}

The values of background to signal ratios and their uncertainties are shown in the last column of the Table~\ref{tab:syst}. In all three experiments the uncertainty of background to signal ratio  is dominated by Li/He background; it is lowest for the Daya Bay due to overburden and strict veto cut (1 $s$) after muon interacting in the volume of anti-neutrino detectors.

\section{Results}

\begin{figure}[tbh]
\vspace*{7.0cm}
\begin{center}
\includegraphics{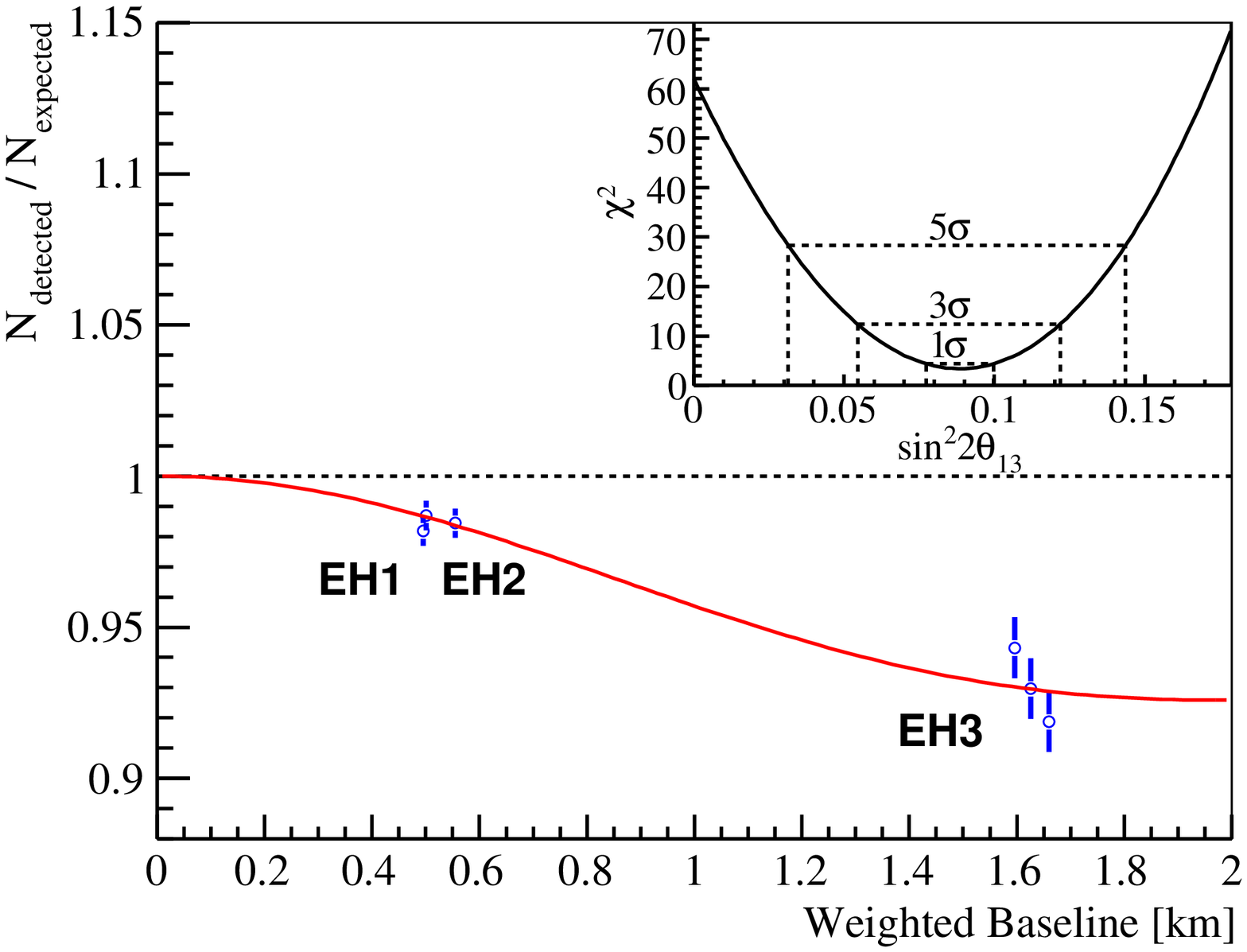}
\includegraphics{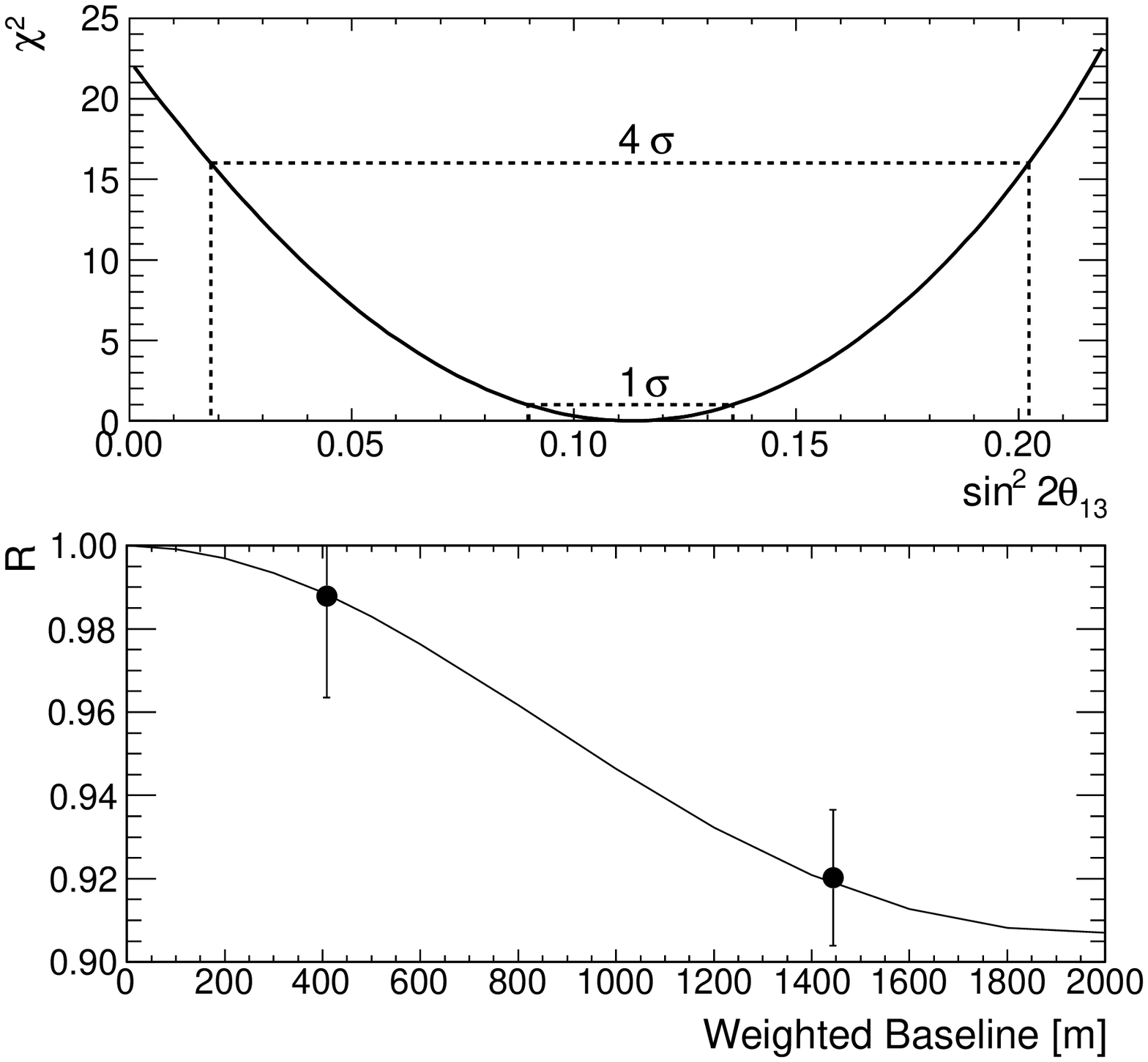}
\caption[*]{ 
Ratio of measured versus expected signals in six Daya Bay detectors (left plot \cite{DYB}) and two RENO detectors (right plot \cite{RENO}). The oscillation survival probability at the best-fit value is given by the smooth curve. The $\chi^2$ value versus $\sin^22\theta_{13}$ are shown in the inset.  Figures are taken from Ref.\cite{DYB,RENO}. \label{fig:s7_chi2} }
\end{center}
\end{figure}
\subsection{Daya Bay}
 The $\bar{\nu}_e$ rate in the far hall was predicted with a weighted combination of the two near hall measurements assuming no oscillation. A ratio of the measured to expected rate is defined as
\begin{eqnarray}
R=\frac{M_f}{\overline{N}_f}=\frac{M_f}{\alpha M_a + \beta M_b}\nonumber ,
\label{eqn:ratio}
\end{eqnarray}
where $\overline{N}_f$ and $M_f$ are the predicted and measured rates in the far hall (sum of AD 4-6), $M_a$ and $M_b$ are the measured, background-subtracted IBD rates in detectors placed in EH1 and EH2, respectively. The values for weights $\alpha$ and $\beta$ were dominated by the baselines, and only slightly dependant on the integrated flux of each core. The ratio observed at the far hall was:
\begin{eqnarray}
R=0.944\pm 0.007({\rm stat.}) \pm 0.003({\rm syst.}) \nonumber ,
\end{eqnarray}
where the statistical (systematic) uncertainties were obtained by propagating statistical (uncorrelated systematic) uncertainties in the measured IBD counts in the three halls.

The value of $\sin^22\theta_{13}$ was determined by minimizing a $\chi^2$ functional constructed with pull terms accounting for the correlation of the systematic errors, see details in \cite{DYB}. 
The best-fit value is
\begin{eqnarray}
\sin^22\theta_{13}=0.089\pm 0.010({\rm stat.})\pm0.005({\rm syst.}) \nonumber ,
\end{eqnarray}
with a $\chi^2$/NDF of 3.4/4. All best estimates of pull parameters are within its one standard deviation based on the corresponding systematic uncertainties. The no-oscillation hypothesis is excluded at 7.7 standard deviations.

\subsection{RENO}
Rate based analysis of data taken by RENO experiment \cite{RENO} yielded following results for ratio of measured to expected rate in the far detector:
\begin{eqnarray}
R_{far} = 0.920 \pm 0.009({\rm stat.}) \pm 0.014({\rm syst.}) \nonumber ,
\end{eqnarray}
and for the value of $\sin^22\theta_{13}$:
\begin{eqnarray}
\sin^2 2\theta_{13} = 0.113 \pm 0.013(\rm stat.) \pm 0.019(\rm syst.) \nonumber ,
\end{eqnarray}
with the significance of 4.9 standard deviations. 

\begin{figure}[htb]
\vspace*{7.0cm}
\begin{center}
\includegraphics{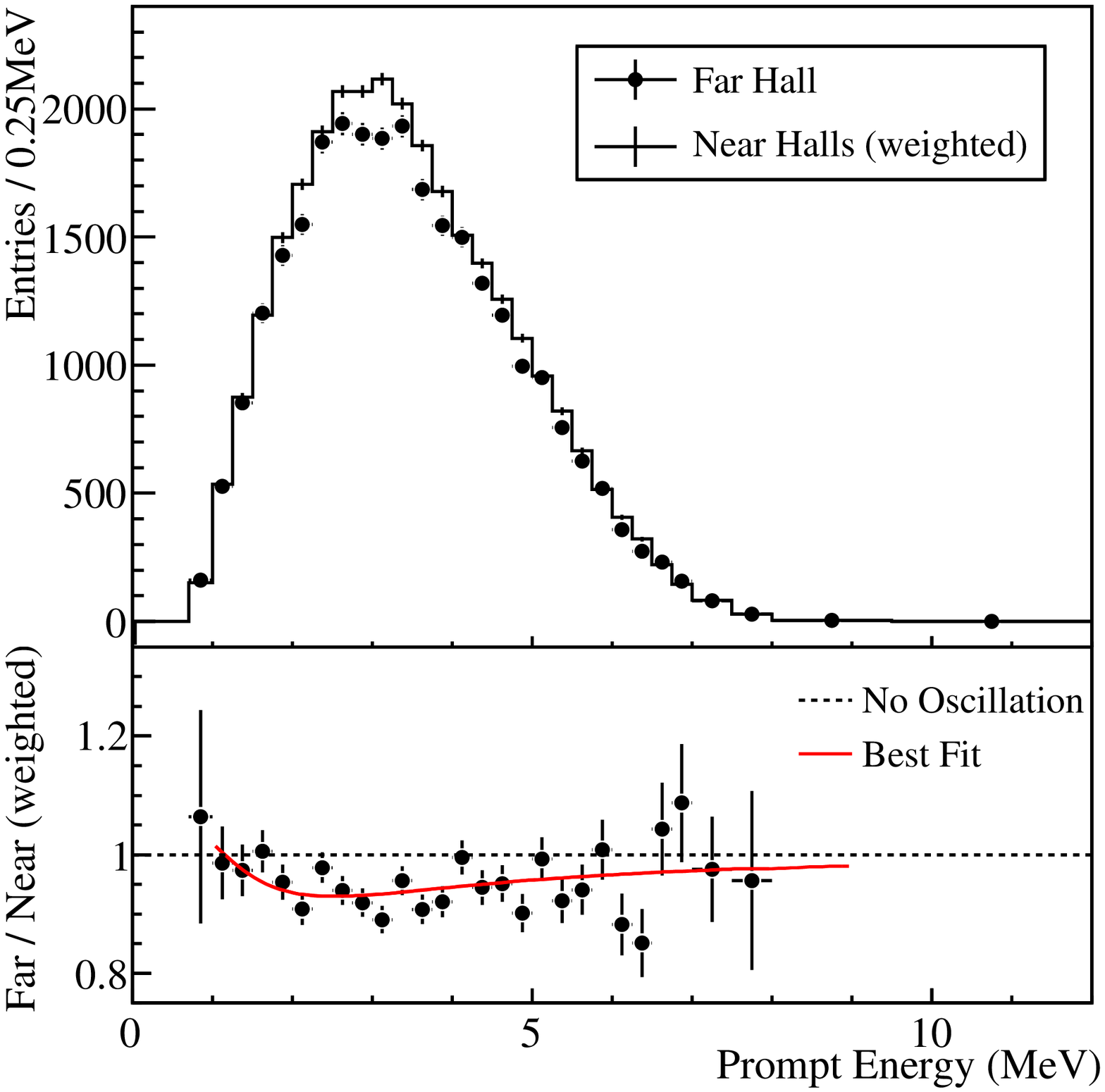}
\includegraphics{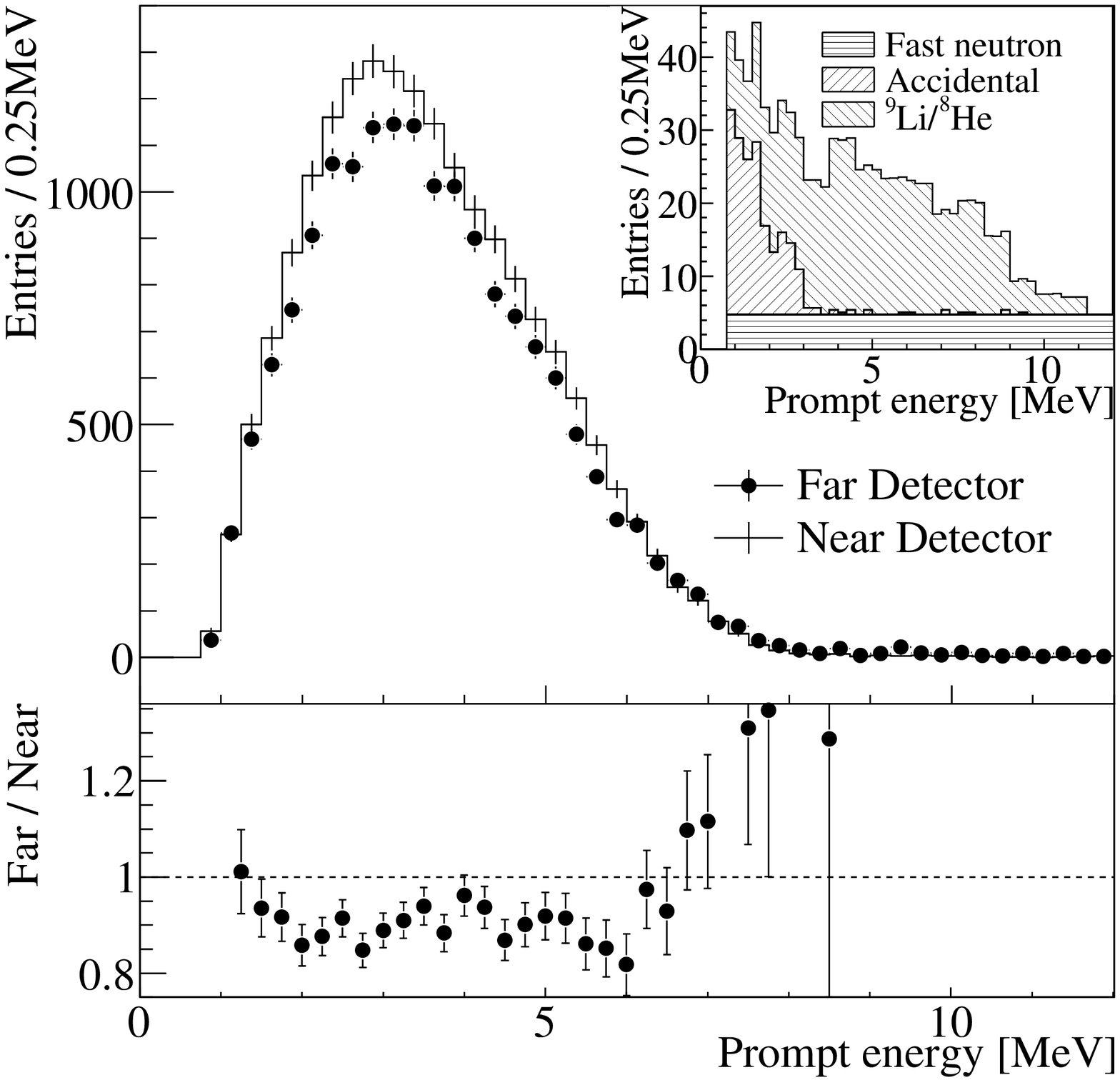}
\caption[*]{ \begin{scriptsize} {\bf Left: Daya Bay}. Top: Measured prompt energy spectrum of the far detectors
compared with the no-oscillation prediction based on the measurements of the near detectors. Bottom: The ratio of measured and predicted no-oscillation spectra. The solid curve is the expected ratio with oscillations, calculated as a function of neutrino energy assuming $\sin^22\theta_{13}=0.089$ obtained from the rate-based analysis. The dashed line is the no-oscillation prediction. 
{\bf Right: RENO}. Top:  Measured prompt energy spectrum of the far detector compared with the no-oscillation prediction based on the measurements of the near detector. Bottom: The ratio of measured and predicted no-oscillation spectra. The dashed line is the no-oscillation prediction. Figures are taken from Ref.~\cite{DYB,RENO}.
 \end{scriptsize}
 \label{fig:s7_rate_deficit}}
\end{center}
\end{figure}

The results of Daya Bay and RENO experiments are illustrated on Figure~\ref{fig:s7_chi2}. Distortion of the observed ratio of far and near $\bar{\nu}_e$ spectra (see Figure~\ref{fig:s7_rate_deficit}) provides further evidence of neutrino oscillations.

\subsection{Double Chooz}
In Double Chooz experiment running without near detector, the data from far detector are compared to calculated (and normalized to Bugey4 data) no-oscillation anti-neutrino spectra. The rate only analysis result\cite{DC2} is:

\begin{eqnarray}
\sin^2 2\theta_{13} = 0.170 \pm 0.035(\rm stat.) \pm 0.040(\rm syst.) \nonumber .
\end{eqnarray}

\begin{figure}[htb!]
\vspace*{8.0cm}
\begin{center}
\includegraphics{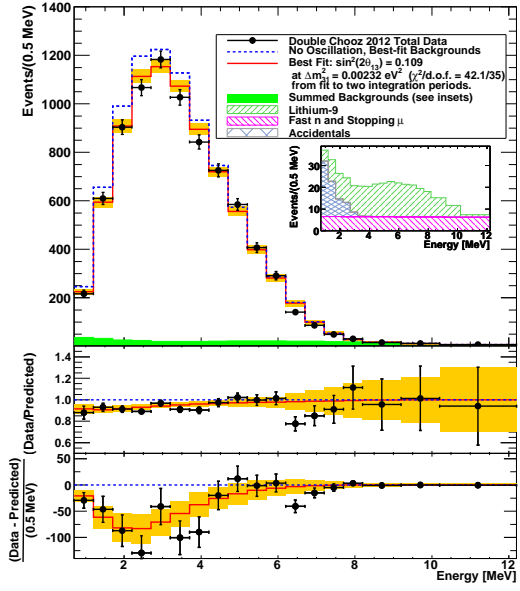}
\caption[*]{ \begin{scriptsize} {\bf Double Chooz}. Top: Measured prompt energy spectrum superimposed on the expected prompt energy spectrum, including backgrounds (green 
region), for the no-oscillation (blue dotted curve) and best-fit (red solid curve) at $\sin^22\theta_{13} = 0.109$.  Inset: 
stacked spectra of backgrounds. Middle: Ratio of data and no-oscillation prediction.  
Bottom: differences between data and no-oscillation prediction 
(data points), 
and differences between best fit prediction and no-oscillation 
prediction (red curve).  The orange band represents the systematic 
uncertainties on the best-fit prediction. Figure is adopted from Ref.~\cite{DC2}. \end{scriptsize}
 \label{fig:DoubleChooz}}
\end{center}
\end{figure}

Double Chooz collaboration reported also combined rate and shape analysis\cite{DC2} result illustrated on Figure~\ref{fig:DoubleChooz}: 
\begin{eqnarray}
\sin^2 2\theta_{13} = 0.109 \pm 0.030(\rm stat.) \pm 0.025(\rm syst.) \nonumber .
\end{eqnarray}

\section{Importance of the result}
Non zero value of $\theta_{13}$ would allow for violation of CP and T symmetry in lepton sector of the Standard Model of elementary particles. 
In particular following CP and T violating differences of appearance probabilities can be measured ($f\neq g$):
\begin{eqnarray}
\Delta^{CP}_{fg}  \left( \frac{x}{E} \right) &\equiv & P_{\nu_{f}\rightarrow \nu_{g}}\left( \frac{x}{E} \right)-P_{\bar{\nu}_{f}\rightarrow \bar{\nu}_{g}}\left( \frac{x}{E} \right) = -\Delta^{CP}_{\bar{f}\bar{g}}  \left( \frac{x}{E} \right)\\
\Delta^{T}_{fg}  \left( \frac{x}{E} \right) &\equiv & P_{\nu_{f}\rightarrow \nu_{g}}\left( \frac{x}{E} \right)-P_{\nu_{g}\rightarrow \nu_{f}}\left( \frac{x}{E} \right) =  -\Delta^{T}_{\bar{f}\bar{g}}  \left( \frac{x}{E} \right) \nonumber
\end{eqnarray}
Due to CPT invariance, they shall have the same values:
\begin{eqnarray}
\Delta^{CP}_{fg}  \left( \frac{x}{E} \right)   & = & \Delta^{T}_{fg}  \left( \frac{x}{E} \right) \equiv \Delta_{fg}  \left( \frac{x}{E} \right) \nonumber
\end{eqnarray}

For three neutrino flavours the variable $\Delta_{fg}$ has the same absolute value for all $f\neq g$ that is given by formula:
\begin{eqnarray}
\Delta  \left( \frac{x}{E} \right) & = & \pm 2 \sin \delta cos \theta_{13} \prod_{i<j} \sin 2\theta_{ij} \sin \left(  \frac{\Delta m^{2}_{ji} x }{4\hbar c E} \right) \nonumber
\end{eqnarray}

Using current measurement of $\theta_{13}$ and other neutrino mixing parameters one can estimate the effect of CP violation to the difference of $P_{\nu_{\mu}\rightarrow \nu_{e}}-P_{\bar{\nu}_{\mu}\rightarrow \bar{\nu}_{e}}$ measured at the first appearance maximum ($L/E = 500~km/GeV$) to be $\approx 3\% \cdot \sin \delta$ .

\section{Summary}
A non zero, surprisingly large value of the third mixing angle $\theta_{13}$ has been measured in 2012. The result is extremely important as it is opening future searches for violation of CP in lepton sector. 

After 2011 hints for non zero value of $\theta_{13}$ from accelerator experiments \cite{T2K,MINOS} and Double Chooz \cite{DC} it is important that today we have consistent results from three different experiments:
\begin{itemize}
\item precise measurement of $\theta_{13}$ by Daya Bay \cite{DYB} experiment with the significance of 7.7 sigma reported at Neutrino 2012; the discovery of non zero value with significance exceeding 5 sigma was announced in March and published in Ref.\cite{DYB1}. 
\item 5 sigma observation of non-zero value of $\theta_{13}$ by RENO announced in April and published in Ref.\cite{RENO}
\item 3 sigma indication of non-zero value of $\theta_{13}$ by Double Chooz \cite{DC2} experiment reported at Neutrino2012. 
\end{itemize}

With more data in near future one can expect improved results with reduced statistical errors and systematic uncertainties. More detailed analysis will include also the shape of detected energy spectra.  

The three experiments collected several hundred thousands of anti-neutrino interactions at different distances from the reactor cores and new interesting analyses will be performed using such unique set of data.


\begin{thebibliography}{0}

\bibitem{DYB1}F.~P.~An {\it et al.} [Daya Bay Collaboration],
  Phys.\ Rev.\ Lett.\  {\bf 108}, 171803 (2012).
  
\bibitem{DYB} F.~P.~An {\it et al.} [Daya Bay Collaboration], Chinese Physics {\bf C37}, 011001 (2013), arXiv:1210.6327 [hep-ex].

\bibitem{RENO}J.~K.~Ahn {\it et al.} [RENO Collaboration],
  Phys.\ Rev.\ Lett.\  {\bf 108}, 191802 (2012)\\
Soo-Bong Kim, Talk at Neutrino 2012, Kyoto, June 4, 2012. 

\bibitem{DC}Y.~Abe {\it et al.}  [Double Chooz Collaboration],
  Phys.\ Rev.\ Lett.\  {\bf 108}, 131801 (2012).
  
\bibitem{DC2}Y.~Abe {\it et al.}  [Double Chooz Collaboration],
  arXiv:1207.6632v4 [hep-ex] 30 Aug 2012 \\
M. Ishitsuka, Talk at Neutrino 2012, Kyoto, June 4, 2012.

\bibitem{ad12} F.~P.~An {\it et al.} [Daya Bay Collaboration], Nucl. Instr. and Meth.\ A {\bf 685}, 78 (2012).  

\bibitem{MIK} L.Mikaelyan and  V.V.Sinev, Phys.Atom.Nucl.{\bf 63}:1002-1006,2000; Yad.Fiz.63 N6:1077-1081,2000. 

\bibitem{T2K}K.~Abe {\it et al.}  [T2K Collaboration],
  Phys.\ Rev.\ Lett.\  {\bf 107}, 041801 (2011).
  
\bibitem{MINOS} P. Adamson et al. [MINOS], Phys. Rev. Lett. {\bf 107}, 181802 (2011).

\end{thebibliography}
\end{document}